\documentclass[preprint, amsmath,amssymb, aps]{revtex4-1}
\usepackage{dcolumn}
\usepackage{amssymb}
\usepackage{subfig}
\usepackage{caption}
\usepackage{float}
\usepackage{graphicx}
\usepackage{epsfig}
\usepackage{amsmath}
\usepackage{float}
\usepackage{epsfig}
\usepackage{hyperref}
\newcommand{\abs}[1]{\left|#1\right|}

\newcommand{\sign}[1]{\text{sgn}\left(#1\right)}
\newcommand{\erf}[1]{\text{Erf}\left(#1\right)}
\newcommand{\erfc}[1]{\text{Erfc}\left(#1\right)}
\begin{document}
\preprint{APS/PRL}
\title{Diffusion-reaction approach to electronic relaxation in solution: Exact time domain solution for Dirac delta function sink model
}
\author{Saravanan Rajendran* and Aniruddha Chakraborty}
\affiliation{School of Basic Sciences, Indian Institute of Technology Mandi, Kamand, HP-175075, India.}
\date{\today}
\begin{abstract}
\noindent We propose an analytical method for finding the time domain solution for the problem of electronic relaxation of a molecule in solution. The relaxation process is modeled by the decay of a diffusing probability distribution through an absorbing sink. In our model, the diffusive motion is modeled using the Smoluchowski equation for harmonic potential and the sink is represented by a Dirac Delta function of arbitrary strength and position. This has been an unsolved problem for a long time and is of immense importance as a model for understanding non-radiative electronic relaxation of a molecule in solution. Our solution can be used to understand various reaction-diffusion problems.
\end{abstract}
\maketitle
\noindent Relaxation of a molecule from an electronically excited state which is immersed in a polar solvent is of immense interest in the past, present, as well as in future, for both experimentalists and theoreticians \cite{Bagchi2,Bagchi,Ben,Lippert}. The molecules initially immersed in the polar solvent is put on an electronically excited state potential by interaction with the appropriate electromagnetic radiation. The interaction from the solvent results in the random changing of configurations in the excited molecule. The random configuration is represented by the corresponding probability distribution function changing over time. 
When the configuration changes, the molecule may undergo non-radiative decay from some part of the potential curve.
 The molecule may also undergo a radiative decay which is assumed to be independent of molecular configurations. The random motion models to the diffusion of a probability distribution 
and de-excitation is represented by the absorption of the molecule. All theoretical models so far assume one dimensional diffusive motion on the excited state potential energy surface and the relevant coordinate is denoted by $x$. It is very standard to assume the diffusive of molecular configuration on the electronically excited state is overdamped and most recent models assume diffusion of molecular configurations happens under the influence of parabolic potential. For theory point of view, the problem is to calculate the probability that the molecule will still remain in the electronically excited state at time $t$. The area under the probability distribution curve is denoted by $Q(t)$ which gives the survival probability of the molecule for being in the electronically excited state.   Therefore the probability $P(x,t)$, that the molecular configuration may be found at $x$ at the time $t$ obeys the following modified Smoluchowski equation given by
\begin{equation}
\frac{\partial P(x, t)}{\partial t} = D\frac{\partial^2 P(x,  t)}{\partial x^2} + k  \frac{\partial}{\partial x} \left( x P(x,  t)\right) - k_{r}P(x,t) - k_{0}S(x)P(x,t),
\label{eqn:harmonicdd1}
\end{equation}
\noindent In the above, $D$ is the diffusion coefficient, $k$ is the force constant for harmonic potential, $S(x)$ is a position dependent sink function taken to be normalized, $k_{0}$ is the strength for non-radiative decay and $k_{r}$ is the rate constant for radiative decay. Before interaction with the incident light, 
the molecule is initially on the ground electronic state and the solvent is at temperature T - the molecular configuration is changing randomly. Then the molecule is excited (Franck-Condon) electronically and therefore $x_0$ is the initial configuration of the excited molecule which is also random. We assume a initial distribution form represented by $P(x_0,0)$. The condition for the population to decay to ground state is represented by a Dirac delta function {\it i.e.}, $S(x) = \delta(x -x_c)$. Now we proceed to find an analytical expression for $Q(t)$, survival probability of the molecule on electronically excited state. It is possible to derive an analytical expression of $Q(t)$ only in one case where $k_0 \rightarrow \infty$ {\it i.e.}, the case of a pin hole sink with the constraint that sink is at the origin\cite{Bagchi,Szabo}. Other solutions of Eq. \ref{eqn:harmonicdd1} was only possible in Laplace domain\cite{Sebastian1} - and it is not possible to invert the solution in time domain. As in almost all the cases decay is never exponential over the entire time scale. Therefore one can define the rate constants, as long time rate constant
\begin{equation}
k_{l}= - \lim_{t \to \infty}\frac{\mathrm{d} \ln Q(t)}{\mathrm{d} t},
\label{eqn:long}
\end{equation}
\noindent and an average rate constant $k_{a}$ by
\begin{equation}
k_{a}^{-1}=\int_{0}^{\infty}Q(t) \mathrm{d}t.
\label{eqn:average}
\end{equation}
\noindent Therefore, time domain solution for the case where sink is not of infinite strength and is not at the origin is an interesting problem to study. In the following we give a simple procedure for finding the exact analytical solution of the problem in time domain. We find it convenient to start with the Eq. \ref{eqn:harmonicdd1} without the Dirac delta function sink term and the corresponding equation is given by
\begin{equation}
\frac{\partial P(x, t)}{\partial t} = D\frac{\partial^2 P(x,  t)}{\partial x^2} + k  \frac{\partial}{\partial x} \left( x P(x,  t)\right),
\label{eqn:harmonic}
\end{equation}
\noindent For Dirac delta function $P(x,0) = \delta (x+x_0)$, the solution of the above equation is known to be \cite{Risken},
\begin{equation}
P(x, t) = \frac{e^{-\frac{(x+x(t))^2}{2\sigma(t)^2}}}{ \sqrt{2\pi}\sigma(t)} ,
\label{eqn:p0}
\end{equation}
\noindent where  $x(t)=x_0e^{-k t}$, $\sigma(t)^2 = \frac{1}{2k}(1 - e^{- 2k t})$. Rearranging the Eq. \ref{eqn:harmonic} gives,
\begin{equation}
\frac{\partial P(x, t)}{\partial t} - k \frac{\partial}{\partial x} \left( x P(x,  t)\right) = D \frac{\partial^2 P(x,  t)}{\partial x^2}.
\label{eqn:3}
\end{equation}
\noindent Now, we insert the $P(x,t)$ given by Eq. \ref{eqn:p0} to compute the L. H. S. of Eq. \ref{eqn:3},
\begin{equation}
\frac{\partial P(x, t)}{\partial t} -  k\frac{\partial}{\partial x} \left( x P(x,  t)\right) = \frac{-Dk(1-e^{-2kt})+k^2x^2}{D(1-e^{-2kt})^2} P(x,t),
\end{equation}
\noindent while substituting the $P(x,t)$ given by Eq. \ref{eqn:p0} gives the R. H. S. of Eq. \ref{eqn:3} as,
\begin{equation}
D\frac{\partial^2 P(x,  t)}{\partial x^2} = \frac{-Dk(1-e^{-2kt})+k^2x^2}{D(1-e^{-2kt})^2} P(x,t).
\end{equation}
\noindent The result is quite obvious. Now we take only the time derivative of $P(x,t)$, and we get
\begin{equation}
 \frac{\partial P(x, t)}{\partial t}  =  e^{-2kt}\frac{-Dk(1-e^{-2kt})+k^2x^2}{D(1-e^{-2kt})^2} P(x,t),
\end{equation}
\noindent By combining the above three equations, the following relation can be written
\begin{equation}
e^{2k t} \frac{\partial P(x, t)}{\partial t}  =  D\frac{\partial^2 P(x,  t)}{\partial x^2}
\end{equation}
\noindent where $P(x,t)$ is given by Eq. \ref{eqn:harmonic}. Now we can re-write the above equation as
\begin{equation}
\frac{\partial P(x,\tau)}{\partial \tau}  =  D \frac{\partial^2 P(x, \tau)}{\partial x^2}
\label{eqn:flat}
\end{equation}
\noindent  Again $P(x,t)$ is given by Eq. \ref{eqn:harmonic},
\begin{equation}
\frac{\partial }{\partial \tau}= \frac{\partial t }{\partial \tau}
\frac{\partial }{\partial t},
\end{equation}
which implies,
\begin{equation}
\frac{\partial t }{\partial \tau} = e^{2 k t},
\end{equation}
and therefore the variable $\tau$ is derived to be
\begin{equation}
\int_{0}^{\tau} d \tau' = \int_{0}^{t} e^{- 2 k t'} dt'  \implies \tau = \frac{1- e^{- 2k t}}{2k}. 
\end{equation}
\noindent The solution of the flat potential yields the solution for harmonic potential by replacing $\tau$ by $\frac{1-e^{-2kt}}{2k}$, and $x_0$ by $x_0e^{-kt}$. The solution of Eq. \ref{eqn:flat} is given by,
\begin{equation}
P(x, \tau) = \frac{e^{ - \frac{(x+x_0)^2}{4 D \tau}}}{ \sqrt{4 \pi D \tau}} ,
\end{equation}
Therefore the solution of Eq. \ref{eqn:harmonic} can be derived to be,
\begin{equation}
P(x, t) = \frac{e^{-\frac{(x+x_0e^{-kt})^2}{2\frac{D}{k}(1-e^{-2kt})}}}{\sqrt{2\frac{D}{k}\pi (1- e^{- 2k t})}} 
\end{equation}
\noindent Now we will add a sink term in Eq.\ref{eqn:harmonic} to get
\begin{equation}
\frac{\partial P(x, t)}{\partial t} = D\frac{\partial^2 P(x,t)}{\partial x^2} +k  \frac{\partial}{\partial x} \left(x P(x,t)\right) - k_r P(x,t) - k_{0}\delta(x-x_c)P(x,t),
\label{eqn:harmonicdd}
\end{equation}
\noindent In terms of new variable $\tau$ the above equation may be written as
\begin{equation}
\frac{\partial P(x, \tau)}{\partial \tau} =D\frac{\partial^2 P(x, \tau)}{\partial x^2}  - k_r P(x,\tau) - k_{0} \delta (x - x_c) P(x,\tau),
\end{equation}
Now we will solve the above equation for {\it i.e.}, for flat potential and in the solution we will use appropriate replacements to obtain the solution of Eq. \ref{eqn:harmonicdd}. The above equation can be solved using the half-Fourier transformation,
\begin{equation}
\tilde P(x,\omega)= \int^\infty_0 P(x,\tau) e^{i \omega \tau} d\tau.
\end{equation}
Laplace transformation of Eq. (1) yields the following equation
\begin{equation}
-i \omega {\tilde P}(x,\omega)-P(x,0)=D \frac{\partial^2{\tilde P}(x,\omega)}{\partial {x}^2} - k_r {\tilde P}(x,\omega) - k_0 \delta(x -{x}_c){\tilde P}(x,\omega).
\end{equation}
\noindent We use $P(x,0) = \delta(x +x_0)$ to derive the spacetime propagator,
\begin{equation}
-i \omega {\tilde P}(x,\omega)-\delta(x+x_0)= D \frac{\partial^2{\tilde P}(x,\omega)}{\partial {x}^2} - k_r {\tilde P}(x,\omega) - k_0 \delta(x -{x}_c){\tilde P}({x}_c,\omega).
\label{eqn:freefourier}
\end{equation}
\noindent Eq. \ref{eqn:freefourier} is now solved using the Green's function method and the solution is given by  
\begin{equation}
{\tilde P}(x,\omega)= \int^\infty_{-\infty} d{x}_{i}{\tilde G}_{0}(x,{x}_i|\omega-ik_r)\delta(x_0 +x_i) - k_0 {\tilde P}({x}_c,\omega)\int^{\infty}_{-\infty} dx_{i}{\tilde G}_{0}(x,x_i|\omega - i k_r)\delta(x_i -{x}_c).
\end{equation}
\noindent On simplification
\begin{equation}
{\tilde P}(x,\omega)={\tilde G}_{0}(x,-x_0|\omega - i k_r) - k_0 {\tilde P}({x}_c,\omega){\tilde G}_{0}(x,x_c|\omega - i k_r).
\end{equation}
\noindent The unknown ${\tilde P}(x_c,\omega)$ can be solved by substituting $x = x_c$ and is obtained to be,
\begin{equation}
{\tilde P}({x}_c,\omega)= \frac{{\tilde G}_{0}(x_c,-x_0|\omega - i k_r)}{1 + k_0 {\tilde G}_{0}(x_c,x_c|\omega - i k_r)}.
\end{equation}
\noindent Now using this we derive the distribution in the Fourier domain,
\begin{equation}
{\tilde P}({x},\omega)={\tilde G}_{0}({x},-x_0|\omega - i k_r) - \frac{ k_0 {\tilde G}_{0}(x,x_c|\omega - i k_r) {\tilde G}_{0}(x_c,-x_0|\omega - i k_r)}{{1 + k_0 {\tilde G}_{0}(x_c,x_c|\omega - i k_r)}}.
\end{equation}
\noindent The Green's function without the Dirac delta sink term is expressed as,
\begin{equation}
{\tilde G}_{0}(x,x_i|\omega - i k_r) = \frac{e^{- \sqrt{\frac{-i \omega + k_r}{D}}|x -x_i |}}{2\sqrt{(- i \omega + k_r)D}},
\end{equation}
which upon substitution yields the following $P(x,\omega)$,
\begin{equation}
{\tilde P}(x,\omega - i k_r)= \frac{e^{- \sqrt{\frac{- i \omega + k_r}{D}}|x +x_0 |}}{2\sqrt{( - i \omega + k_r)D}}   -\frac{k_0 e^{-\sqrt{\frac{ - i \omega + k_r}{D}}(|x + x_0|+|x - x_c|)}}{2\sqrt{(- i \omega + k_r) D}(2 \sqrt{ (- i \omega + k_r) D}+k_0)}.
\end{equation}
\noindent The inverse transformation leads to an analytical expression for $P(x, t)$ as given by,
\begin{equation}
   P(x,\tau)= \frac{e^{-\frac{(x+ x_0)^2}{4 D c} - k_r \tau}}{2 \sqrt{\pi D \tau}}-\frac{k_0}{4D}e^{\frac{k_0^2}{4D}\tau-k_r\tau+\frac{k_0}{2D}(\abs{x-x_c}+\abs{x_c+x_0})}\erfc{\frac{k_0}{2\sqrt{D}}\sqrt{\tau}+\frac{\abs{x-x_c}+\abs{x_c+x_0}}{2\sqrt{D\tau}}}.
\end{equation}
The survival probability can be obtained by integrating the distribution over all x and is given as\cite{Saravanan},
\begin{eqnarray}
Q(t)=e^{-k_r\tau}[1+e^{\epsilon_0(x_0+x_c)+\epsilon_0^2D\tau}\left(\sign{\frac{x_0+x_c}{2D\tau}+\epsilon_0}-\erf{\frac{x_0+x_c+2\epsilon_0D\tau}{2\sqrt{Dt}}}\right)\\ \nonumber+2\theta(-x_0-x_c-2\epsilon_0D\tau)e^{-\epsilon_0^2D\tau}-\erfc{\frac{x_0+x_c}{\sqrt{4D\tau}}}].
\label{eqn:pt}
\end{eqnarray}
With replacements, $\epsilon_0=\frac{k_0}{2D}$, $\tau=\frac{1- e^{- 2k t}}{2k}$, and $x_0=x_0e^{-kt}$, the above solution yields the solution of harmonic potential with $\delta$-function sink. Using the $Q(t)$ expression, one can calculate long time and average rate constants given by expressions \eqref{eqn:long} and \eqref{eqn:average}.
\section{Results and Discussion}
\subsection{Analytical and Numerical verifications}
\begin{figure}
    \centering
    \subfloat[][]{\includegraphics{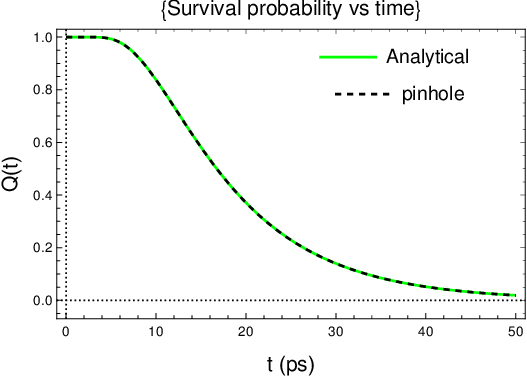}\label{fig:0a}}\subfloat[][]{\includegraphics{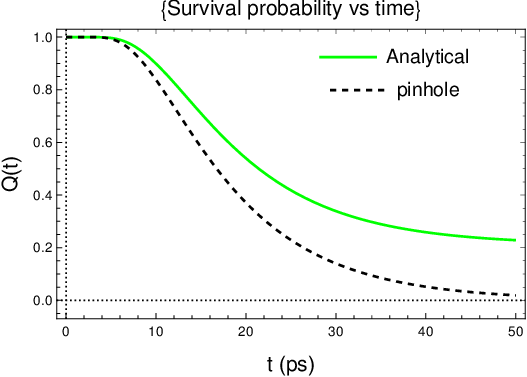}\label{fig:0b}}
    \caption{Comparison with the asymptotic limit $k_0\to\infty$ for the following parameters: $D = 0.2$ nm$^2$/ps, $x_0 = 5$ nm, $x_c = 0$ nm, $k = .1$ ps$^{-1}$, $k_r = 0$ ps$^{-1}$, \protect\subref{fig:0a} $k_0 = 100$ nm/ps, \protect\subref{fig:0b} $k_0 = 1$ nm/ps. It is to note that for smaller values of $k_0$, the pinhole result and the analytical result does not match.}
    \label{fig:pinholecomps}
\end{figure}

The presented analytical solution is verified under the asymptotic limit $k_0\to\infty$ and $x_c=0$. In this limit, the problem reduces to the pinhole sink case placed at origin. The Fig. \ref{fig:pinholecomps} compares the presented result (Eq. \ref{eqn:pt}) with analytic formula obtained in Ref. \cite{Bagchi} for the pinhole sink case. For the finite strength case, the solution is verified against the numerical solver NDSolve of Mathematica. The $\delta$-function is taken to be a Gaussian function with $\sigma=5\times 10^{-4}$. The analytical result agrees with the numerical calculations.
\begin{figure}    \centering    \includegraphics[scale=1]{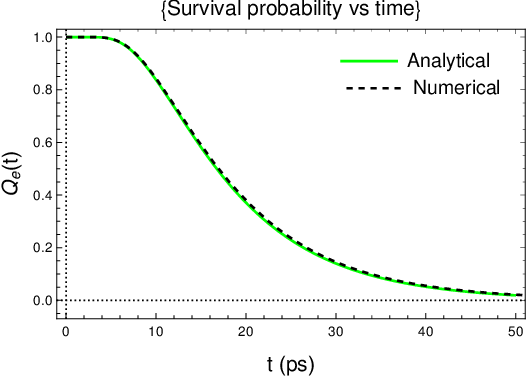}    \caption{Comparison of the analytical survival probability $Q(t)$ with the numerical solution for the following parameters: $D = 0.2$ nm$^2$/ps, $k_0$ = 3 nm/ps, $x_0$ = 5 nm, $x_c$ = 0 nm, $k = .1$ ps$^{-1}$, $k_r$ = 0 ps$^{-1}$. The $\delta$-function is mimicked by a normalized Gaussian sink with a small width given by, $\sigma = 5\times 10^{-4}$ nm .}    \label{fig:numcomp}\end{figure}
\subsection{Theory of electronic relaxation in solutions}
\begin{figure}
    \centering
    \includegraphics[scale=1]{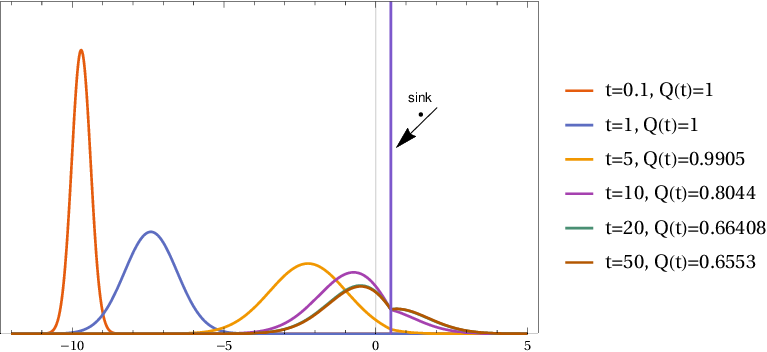}
    \caption{The plot of the distribution $P(x,t)$ for different time presented along with the area under the distribution curve. The parameters are: $D = 0.5$ nm$^2$/ps, $k_0$ = 1 nm/ps, $x_0$ = 10 nm, $x_c$ = 0.5 nm, k = .3 ps$^{-1}$, $k_r$ = 0 ps$^{-1}$.}
    \label{fig:fate}
\end{figure}

\begin{figure}
    \centering
    \subfloat[][]{\includegraphics{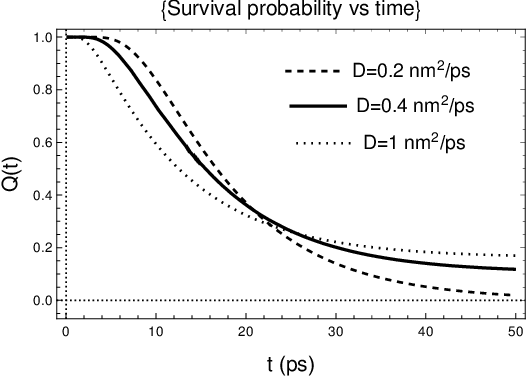}\label{fig:1a}}\subfloat[][]{\includegraphics{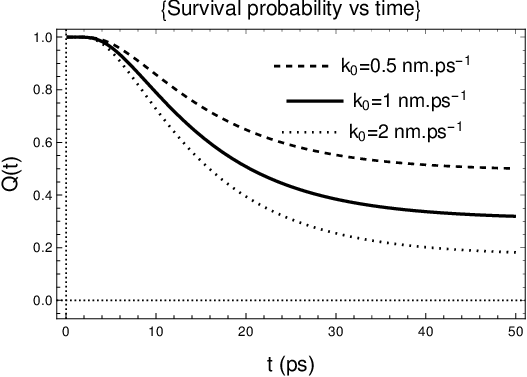}\label{fig:1b}}
    \caption{The survival probability $Q(t)$ versus time is plotted showing\protect\subref{fig:1a} the effect of Diffusivity D while other parameters are: $k_0 = 3$ nm/ps, $x_0 = 5$ nm, $x_c = 0$ nm, $k = .1$ ps$^{-1}$, $k_r = 0$ ps$^{-1}$, \protect\subref{fig:1b}the effect of nonradiative decay strength $k_0$ while other parameters are: $D=0.5$ nm$^2$/ps, $x_0 = 5$ nm, $x_c = 0$ nm, $k = .2$ ps$^{-1}$, $k_r = 0$ ps$^{-1}$ .}
    \label{fig:eff1}
\end{figure}

\begin{figure}
     \centering
    \subfloat[][]{\includegraphics{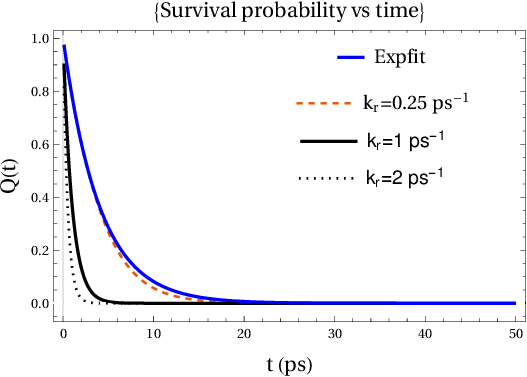}\label{fig:2a}}\subfloat[][]{\includegraphics{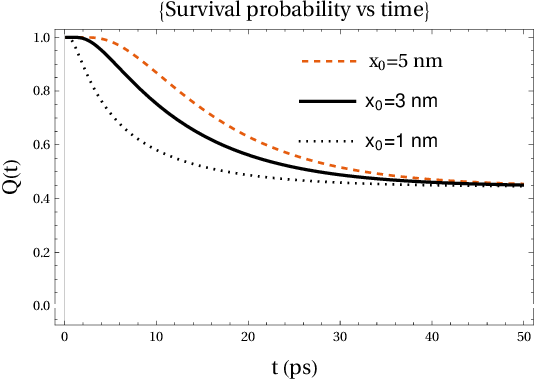} \label{fig:2b}}
    \caption{The survival probability $Q(t)$ versus time plot showing\protect\subref{fig:2a} the effect of radiative decay rate ($k_r$) when the parameters are: $(D, k, k_0, x_c, k_r, x_0)$=(0.5, 0.2, 3, 0, $k_r$, 5), \protect\subref{fig:2b} the effect of initial position of the walker ($x_0$) when the parameters are: $(D, k, k_0, x_c, k_r, x_0)$=(0.5, 0.1, 3, 1, 0, $x_0$).}
    \label{fig:eff2}
\end{figure}

\begin{figure}
     \centering
    \subfloat[][]{\includegraphics{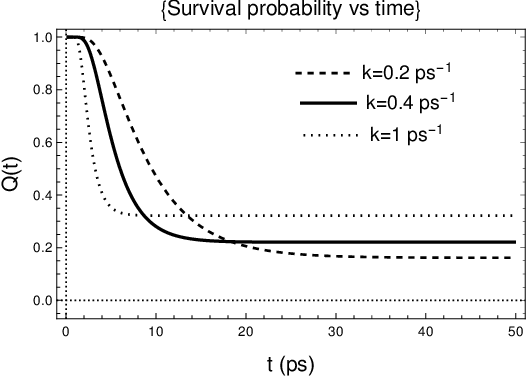}\label{fig:3a}}\subfloat[][]{\includegraphics{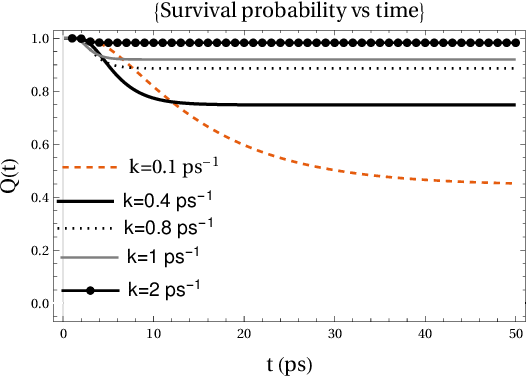} \label{fig:3b}}
    \caption{The survival probability $Q(t)$ versus time plot showing\protect\subref{fig:3a}the effect of slope of the potential when the reactive motion is barrierless, i.e., the reactive channel is at the potential minimum ($x_c=0$) when the other parameters are: $(D, k, k_0, k_r, x_0)$=(0.5, $k$, 3, 0, 5),  \protect\subref{fig:3b} the effect of the slope in the presence of a barrier. The parameters are: $(D, k, k_0, x_c, k_r, x_0)$=(0.5, k, 3, 1, 0, 4).}
    \label{fig:eff3}
\end{figure}

\begin{figure}
     \centering
    \subfloat[][]{\includegraphics{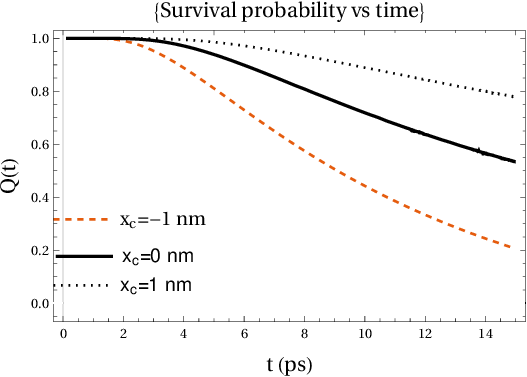}\label{fig:4a}}\subfloat[][]{\includegraphics{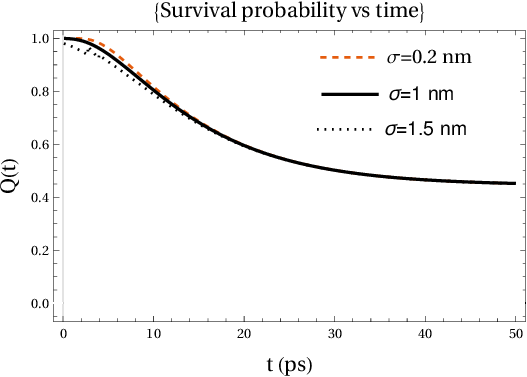} \label{fig:4b}}
    \caption{TThe survival probability $Q(t)$ versus time plot showing \protect\subref{fig:4a}the effect of the distance of the sink from the minimum position. The other parameters are: $(D, k, k_0, k_r, x_0, x_c)$=(0.5, 0.2, 3, 0, 5, $x_c$),  \protect\subref{fig:4b} the effect of $\sigma$. The parameters are: $(D, k, k_0, x_c, k_r, x_0)$=(0.5, 0.2, 3, 1, 0, 4).}
    \label{fig:eff4}
\end{figure}

The Fig. \ref{fig:fate} depicts the dynamics of the distribution on a parabolic potential in the presence of an absorbing boundary. The walker starts at the point $x=-x_0$ at $t=0$, takes a random walk while the mean position of the walker undergoes a translational motion towards the minimum of the potential. The action of diffusion widens the probability distribution. Once the particles reach the potential minimum, there are no translational motion and only diffusion occurs with a scaled diffusion coefficient. The particles face a potential barrier, to attain the reactive configuration. At the absorbing boundary $x=x_c$, the distribution develops a cusp which is required by the flux discontinuity requirement. Once the diffusing distribution meets the sink, the area of the distribution starts reducing over time. The area in the electronic relaxation sense, is the number of molecules surviving on the excited state till time $t$ and is denoted by $Q(t)$. With this picture of the distribution dynamics, we study the effect of system parameters on the survival probability of the random walker. In the diffusion-reaction approach for condensed-phase reactions, the solvent parameters are incorporated through the noise that gives rise to the second-order diffusive term. The diffusivity is of the form, $D=\frac{k_BT}{\zeta}$ where $\zeta$ is the friction coefficient and $T$ is the temperature of the solvent. When the diffusivity is more, the tail of the distribution could reach the sink even for smaller times, and have a faster decay. For the selected set of parameters in Fig. \ref{fig:1a}, the mean distribution reaches the sink typically at $t \sim 25$ ps. After they meet the sink, the distribution with lesser diffusivity spends more time near the sink, thus increasing the probability of decay. The flux leaving the state is characterized by the term, $-k_0P(x_c,t)$ meaning that a sharper distribution at $x=x_c$ will have more decay at that instant. The strength of the boundary can be related to molecular terms as the non-adiabatic coupling strength between the molecular potential surfaces. Higher the coupling strength, higher the non-radiative decay constant $k_0$. When the absorbing boundary has high strength $k_0$, the decay is higher. The non-radiative decay of the excited molecule is localized to a particular configuration. On the contrary, the radiative decay occurs anywhere from the potential surface, hence a slightly bigger $k_r$, can dominate the radiative-nonradiative decay competition (Fig. \ref{fig:2a}). For a radiative decay rate, $k_r\gtrsim 0.25$ ps$^{-1}$, the population profile can be fitted with the single exponential decay. The effect of excitation wavelength appears through the initial position of the excited state molecules. The lower the excitation wavelength, far the peak distribution is placed from the potential minima. Fig. \ref{fig:2b} shows that the distribution that is far from the relaxing configuration, undergoes a slower decay. The effect of the spring constant along the reaction coordinate is shown in Fig. \ref{fig:3a}. The barrierless reaction is one, when the initially prepared reactant species decays without facing any barrier. When the sink is placed at the potential minima, \textit{i.e.}, $x_c=0$, more the `$k$' lead to a faster decay at initial times (Fig. \ref{fig:3a}). For longer time, only a potential with less depth allows more time for different parts of the distribution to interact with the sink. This leads to a situation that overall decay will be more for a lesser $k$. When the reactive configuration is to be attained by climbing a barrier (\textit{i.e.,} $x_c>0$), a more slope would mean less decay at all times. The position of the absorbing boundary determine 3 cases of reaction propagation, $x_c< 0$ or $x_c=0$ would mean a barrierless process, with the distribution facing/not facing a translational motion at the boundary respectively. The boundary when placed closer to the distribution leads to faster decay as expected (Fig. \ref{fig:4a}). For the case $x_c>0$, the relaxation process faces a potential barrier, and hence, a lesser decay to ground state. Another interesting excitation parameter, is the bandwidth of the source. When the incident light is not of monochromatic character, the dye molecules does not excite to a particular bond length. While most of the molecules excite to a particular bond length, some excite to nearby configurations. So if the initial distribution is assumed to take a Gaussian form as given by,
\begin{eqnarray}
P(x,0)=\frac{1}{2\sqrt{\pi\sigma^2}}e^{-\frac{(x+x_0)^2}{4\sigma^2}}.
\end{eqnarray}
Fig. \ref{fig:4b} shows the effect of $\sigma$ on the electronic relaxation process. It is seen that for smaller times, the distribution with a larger $\sigma$, decays faster. The effect of $\sigma$ appears as a correction term $\sigma^2e^{-2kt}$ in addition to the diffusion term. But in the longer run, the correction term $\sigma^2e^{-2kt}$ vanishes and hence there is no effect of $\sigma$ on the decay.
\section{Conclusions}
\noindent We give a time domain solution for the problem of radiation less decay modeled by a Smoluchowski equation equation, with a Dirac delta function sink. The analytical expression $P(x,t)$ obeys boundary conditions and $Q(t)$ is verified against numerical results and with the asymptotic analytical results. The work presents the time-dependent concentration profile for reactions in condensed phase. The profile $Q(t)$ can be useful to experimentalists for fitting the chemical kinetic data as a function of system, and molecular parameters. 

\end{document}